\documentstyle[sprocl]{article}

\input{psfig.sty}

\def\Journal#1#2#3#4{{#1} {\bf #2}, #3 (#4)}

\def\NPB{{Nucl. Phys.} B}
\def\PLB{{Phys. Lett.}  B}
\def\PRL{Phys. Rev. Lett.}
\def\PRD{{Phys. Rev.} D}

\def\PRP{Phys. Rep.}

\begin{document}
\setlength{\unitlength}{1mm}

\title{CHIRAL SYMMETRY IN SUPERSYMMETRIC THREE DIMENSIONAL QUANTUM 
ELECTRODYNAMICS}

\author{M. L. WALKER and C. J. BURDEN}

\address{Department of Theoretical Physics,\\ Research School of 
Physical Sciences and 
Engineering,\\ Australian National University, Canberra, A.C.T. 0200
Australia\\E-mail: mlw105@rsphysse.anu.edu.au, conrad.burden@anu.edu.au}

\maketitle

\begin{abstract}
We describe the investigation 
of spontaneous mass-generation and chiral symmetry breaking in 
supersymmetric QED3 using numerical solutions of the Dyson-Schwinger 
equation together with the CJT effective action and supersymmetric Ward 
identities.  We find that, within the quenched, bare vertex approximation, 
the chirally symmetric solution is favoured.  
\end{abstract}

\setcounter{equation}{0}
\section{Introduction} \label{intro}

Any realistic attempt to unify gravity with the other fundamental
forces of nature must almost certainly be supersymmetric. Whilst perturbative
results are often easier to find in supersymmetric field theories than in
ordinary gauge theories, especially with the use of superfields, the 
development of non-perturbative tools for supersymmetric theories has not 
yet reached maturity.  In this 
paper we apply a nonperturbative tool which has proved to be effective 
in the study of strong interactions, namely Dyson-Schwinger equations 
(DSEs)~\cite{RW94}, to one of the simplest supersymmetric gauge theories.  

We choose as our gauge theory supersymmetric quantum electrodynamics in 
three space-time dimensions (SQED3).  
In its nonsupersymmetric form, three dimensional quantum electrodynamics 
(QED3) with four component spinors~\cite{Pski} exhibits the interesting 
nonperturbative phenomena of confinement and, at least for small numbers 
of flavours, chiral symmetry breaking.  Much of what is known about the 
behaviour of QED3 has been learnt from numerical DSE~\cite{M96} or 
lattice~\cite{DKK90} calculations.  Considering the success of DSE 
techniques in nonsupersymmetric theories, there is a surprising scarcity of 
applications of the method to supersymmetric theories in the literature, 
and an almost total absence of numerical DSE calculations in particular.  

The version of SQED3 which we consider is the four-component 
fermion version first proposed by Pisarski~\cite{Pski}, who 
approached the model by dimensional reduction from SQED4.  Herein, we 
develop SQED3 using the Wess Zumino construction~\cite{wz}.  Our 
construction produces supersymmetry multiplets 
which differ slightly from those of Pisarski because of an extra degree 
of freedom in the charge conjugation matrix.  Pisarski's approach was 
an analytic one based on a $1/N_{\rm flavor}$ expansion.  His analysis 
indicates the existence of a dynamical mass generating solution in the large 
$N_{\rm flavor}$ limit.  

Koopmans and Steringa~\cite{koops} have analysed SQED3 with 2-component
fermions using DSEs 
and claim the critical number $N_c = 3.24$ of 
flavours below which spontaneous mass generation occurs.  This result 
should be treated with some caution however as it ignores the possibility 
of spontaneous parity violation~\cite{A86} and the generation of a photon 
mass via a Chern-Simons term when the number of two-component fermions 
is odd.  Furthermore, there is no indication whether spontaneous mass 
generation corresponds to the stable vacuum of the theory.  

The above treatments use a component formalism to study SQED3.  By contrast, 
Clark and Love~\cite{Love} develop DSEs for SQED4 using a superfield 
formalism.  While more elegant as a mathematical formalism, the superfield 
approach has the disadvantage that each DSE contains an infinite 
number of terms.  This is dealt with by truncating from the chiral 
multiplet self energy diagrams containing seagull and higher order 
$n$-point vertices.  They find that the effective mass contains a prefactor 
which vanishes in Feynman gauge and conclude that there can be no 
spontaneous mass generation in SQED4.  

The work of Clark and Love has been criticised by 
Kaiser and Selipsky \cite{kaiser} on two grounds.  Firstly they argue that 
the truncation of seagull diagrams is too severe as it ignores contributions 
even at the one-loop level.  Secondly they point out that infinities 
arising from infrared divergences which plague the superfield formalism 
can counter the vanishing prefactor and allow spontaneous mass generation.  
These criticisms highlight some of the dangers of attempting to extract 
phenomenological consequences of supersymmetric DSEs by working 
solely with the superfield formalism.  In fact, analyses in the literature of 
SQCD~\cite{sham}, have generally found the component formalism to be the 
most efficient way to proceed.  

In this paper we present a first detailed numerical analysis of the 
chiral multiplet DSE for a supersymmetric gauge theory.  We work with 
the component formalism rather than the superfield formalism to avoid 
the problems encountered by Clark and Love.  By choosing to work in three 
dimensions, and with the component formalism, we find that all integrals 
encountered are infrared and ultraviolet convergent.  

Section \ref{algebra} describes the four-component fermion Clifford 
algebra of QED3 and gives the supersymmetry multiplets.  In section \ref{DSE}
we present the SQED3 lagrangian and chiral multiplet propagator DSEs. 
We explain how supersymmetric Ward Identities can be used to simplify what 
would otherwise be an arduous problem and present our numerical solutions
for quenched SQED3 in the bare vertex approximation. 
We present both chirally symmetric and
asymmetric solutions and use the Cornwall-Jackiw-Tomboulis (CJT) effective 
potential to determine that the massless solution is dynamically favoured.

\setcounter{equation}{0}
\section{The algebra of SQED3} \label{algebra}

In 2+1 dimensions there are two inequivalent, irreducible representations of
the Clifford algebra, given in terms of $2 \times 2$ matrices. These two
representations differ by a minus sign \cite{Sohn}, and have the undesirable
property that either one leads to a version of QED3 which is parity
non-invariant. To circumvent this property, it is common to consider a four-component version of QED3 incorporating Dirac matrices which are a direct sum of the two inequivalent representations \cite{Pski}. The Dirac matrix algebra we
employ here is constructed as follows \cite{burd}:

The $4 \times 4$ matrices $\gamma_\mu$ satisfy $\{ \gamma_\mu,\gamma_\nu \} =
2\eta_{\mu \nu}$, $\eta_{\mu \nu} = \mbox{diag}(1, -1, -1)$ where $\mu$ takes
 the values 0,1 and 2. We take the complete set of 16 matrices
\begin{displaymath}
\{ \gamma_A \} = \{ I, \gamma_4, \gamma_5, \gamma_{45}, \gamma_\mu, 
\gamma_{\mu 4}, \gamma_{\mu 5}, \gamma_{\mu 45} \},
\end{displaymath}
\begin{displaymath}
\gamma_0 = \left( \begin{array}{cc}
\sigma_3 & 0 \\
0 & -\sigma_3 \end{array} \right), \;
\gamma_{1,2} = -i \left( \begin{array}{cc}
\sigma_{1,2} & 0 \\
0 & -\sigma_{1,2} \end{array} \right),
\end{displaymath}
\begin{displaymath}
\gamma_4 = \gamma^4 = \left( \begin{array}{cc}
0 & I \\
I & 0 \end{array} \right), \;
\gamma_5 = \gamma^5 = \left( \begin{array}{cc}
0 & -iI \\
iI & 0 \end{array} \right), \;
\gamma_{45} = \gamma^{45} = -i\gamma_4 \gamma_5 ,
\end{displaymath}
\begin{displaymath}
\gamma_{\mu 4} = i\gamma_\mu \gamma_4, \;
\gamma_{\mu 5} = i\gamma_\mu \gamma_5, \;
\gamma_{\mu 45} = \gamma_\mu \gamma_{45}. 
\end{displaymath}
The matrices $I, \gamma_4, \gamma_5, \gamma_{45}$ are the Pauli matrices in 
block form and as such generate a $U(2)$ algebra. The parity and charge 
conjugation rules for four-component Dirac spinors are given by
\begin{eqnarray}
{\cal P} \psi(x) {\cal P}^{-1} = \Pi \psi(x^0,-x^1,x^2), \; & & 
{\cal P} \bar{\psi}(x) {\cal P}^{-1} = \bar{\psi}(x^0,-x^1,x^2) \Pi^{-1}, 
                                  \nonumber \\
{\cal C} \psi {\cal C}^{-1} = C\bar{\psi}^T, \; & &
{\cal C} \bar{\psi} {\cal C}^{-1} = -\psi^T C^{-1}, 
\end{eqnarray}
where
\begin{equation}
\Pi = \gamma_{14} \mbox{e}^{i\phi_P \gamma_{45}}, \; 
  C = \gamma_2 \mbox{e}^{i\phi_C \gamma_{45}},
\end{equation}
and $(0 \leq \phi_P, \phi_C < 2\pi)$. A Majorana spinor is one for which
$\psi = C\bar{\psi}^T$.
The arbitrary phases $\phi_C$ and $\phi_P$ are important for classifying the bound states in QED3 \cite{burd}.

We have
\begin{equation} \label{cpconj} \begin{array}{c}
C^{-1} \left( \begin{array}{c} \gamma_4 \\ \gamma_5 \end{array} \right) C =
R_C \left( \begin{array}{c} \gamma_4^T \\ \gamma_5^T \end{array} \right),
\\
\\
\Pi^{-1} \left( \begin{array}{c} \gamma_4 \\ \gamma_5 \end{array} \right) \Pi
= R_P \left( \begin{array}{c} \gamma_4 \\ \gamma_5 \end{array} \right),
\end{array} \end{equation}
where
\begin{equation} \label{RpRc} \begin{array}{cc}
R_P = \left( \begin{array}{cc}
-\cos 2\phi_P & -\sin 2\phi_P \\
-\sin 2\phi_P & \cos 2\phi_P
\end{array} \right)
,&
R_C = \left( \begin{array}{cc}
-\cos 2\phi_C & \sin 2\phi_C \\
\sin 2\phi_C & \cos 2\phi_C
\end{array} \right).
\end{array} \end{equation}

The first step in the extension of QED3 to supersymmetry is the construction of chiral multiplets. A reasonable tentative suggestion is 
\begin{equation}
\begin{array}{l}
\delta a = -i\bar{\zeta} \psi \\
\delta b = \bar{\zeta} \gamma_5 \psi \\
\delta \psi = (f+i\gamma_5 g) + i\not \! \partial (a+i\gamma_5 b) \zeta \\
\delta f = \bar{\zeta} \not \! \partial \psi \\
\delta g = i\bar{\zeta} \gamma_5 \not \! \partial \psi,
\end{array}
\end{equation}
where $a$, $b$, $f$ and $g$ are real and $\zeta$ and $\psi$ are Majorana.
If the superalgebra is to hold then the commutator of two of these 
transformations must obey 
\begin{equation}
[ \, \delta_1 , \delta_2 \, ] X = 
2\bar{\zeta}_2 \gamma^\mu \zeta_1 \partial_\mu X, \label{susyalg}
\end{equation}
where X is any component of the multiplet. For the case of $a$ we have
\begin{displaymath}
[ \, \delta_1, \delta_2 \,] a = \bar{\zeta}_2 \gamma^\mu \zeta_1 \partial_\mu a
-i\bar{\zeta}_2 \gamma_5 \gamma^\mu \zeta_1 \partial_\mu b 
- (1 \longleftrightarrow 2).
\end{displaymath}
Now, since $\zeta$ is Majorana
\begin{equation} \begin{array}{lcc}
\bar{\zeta}_1 \gamma^\mu \zeta_2 & = & - \bar{\zeta}_2 \gamma^\mu \zeta_1,
\end{array} \end{equation}
as required, but using equations (\ref{cpconj}) and (\ref{RpRc})
\begin{equation} \begin{array}{lcl}
-i\bar{\zeta}_1 \gamma_5 \gamma^\mu \zeta_2
& = & \bar{\zeta}_2 (\gamma_5 \cos 2\phi_C + \gamma_4 \sin 2\phi_C ) \gamma^\mu
\zeta_1 ,
\end{array} \end{equation}
which does not cancel $\bar{\zeta}_2 \gamma_5 \gamma^\mu \zeta_1$. A similar 
situation arises if $a$ is replaced by any other member of the multiplet. To 
stop this `blowing out' of terms we could simply set $\phi_C = 0$. This would 
be unfortunate though as the angular freedom $\phi_C$ in the matrix C would be 
lost. As stated earlier, this angle is important for classifying the bound 
states in QED3 so we would like to preserve it if possible to
see what effects it has (if any) in the supersymmetric theory.
To this end we define the rotated Dirac matrices.
This can be done by making the substitution
\begin{equation} \label{pwsubst}
\left( \begin{array}{c} \gamma_4 \\ \gamma_5 \end{array} \right)
\longrightarrow
\left( \begin{array}{c} \gamma_P \\ \gamma_W \end{array} \right) =
\left( \begin{array}{cc}
\cos \phi_C & -\sin \phi_C \\
\sin \phi_C & \cos \phi_C
\end{array} \right)
\left( \begin{array}{c} \gamma_4 \\ \gamma_5 \end{array} \right) =
M \left( \begin{array}{c} \gamma_4 \\ \gamma_5 \end{array} \right),
\end{equation}
in the Clifford algebra. (Note that $-i\gamma_P \gamma_W = \gamma_{45}$ so the
matrices $I, \gamma_P, \gamma_W, \gamma_{45}$ again generate an $SU(2)$ 
algebra.) Then
\begin{equation}
C^{-1} \left( \begin{array}{c} \gamma_P \\ \gamma_W \end{array} \right) C =
\left( \begin{array}{c} -\gamma^T_P \\ \gamma^T_W \end{array} \right),
\end{equation}
since
\begin{equation}
M R_C M^{-1} = \left( \begin{array}{cc} -1 & 0 \\ 0 & 1 \end{array} \right).
\end{equation}
The rotated matrices can be used to define a supersymmetry transformation 
consistent with Eq.~(\ref{susyalg}).  In order to do this, all terms except 
$\bar{\zeta}_1 \gamma^\mu \zeta_2$ generated by $\delta_1 \delta_2 X$ must 
be symmetric under interchange of $\zeta_1 , \zeta_2$, i.e.
\begin{equation}
\bar{\zeta}_2 \gamma_W \gamma^\mu \zeta_1 = \bar{\zeta}_1 \gamma_W \gamma^\mu
\zeta_2.
\end{equation}
In this sense $\gamma_W$ is well-behaved but $\gamma_P$ and $\gamma_{45}$
are problem matrices because
\begin{equation}
\bar{\zeta}_2 (\gamma_P, \gamma_{45}) \gamma^\mu \zeta_1
= -\bar{\zeta}_1 (\gamma_P, \gamma_{45}) \gamma^\mu \zeta_2.
\end{equation}
Making the substitution (\ref{pwsubst}) gives the 2+1 dimensional chiral 
multiplet
\begin{equation} \label{Phitrans}
\begin{array}{l}
\delta a = -i\bar{\zeta} \psi \\
\delta b = \bar{\zeta} \gamma_W \psi \\
\delta \psi = (f+i\gamma_W g)+i\gamma^\mu \partial_\mu (a+i\gamma_W b) \zeta \\
\delta f = \bar{\zeta} \not \! \partial \psi \\
\delta g = i\bar{\zeta} \gamma_W \not \! \partial \psi,
\end{array}
\end{equation}
which is the analogue of the standard chiral multiplet in $3+1$ dimensions.

The relative difference between the arbitrary phases $\phi_P$ and $\phi_C$ is
fixed by the imposition of supersymmetry.
Indeed, from equations (\ref{cpconj}) and (\ref{pwsubst}) we have that
\begin{displaymath}
\Pi^{-1} \left( \begin{array}{c} \gamma_P \\ \gamma_W \end{array} \right) \Pi
= M R_P M^{-1} \left( \begin{array}{c}
\gamma_P \\ \gamma_W \end{array} \right)
\end{displaymath}
where
\begin{equation}
M R_P M^{-1} = \left( \begin{array}{cc}
-\cos 2(\phi_P + \phi_C) & -\sin 2(\phi_P + \phi_C) \\
-\sin 2(\phi_P + \phi_C) & \cos 2(\phi_P + \phi_C)
\end{array} \right).
\end{equation}
If the form of the chiral multiplet transformation Eq.~(\ref{Phitrans}) 
is to be maintained under parity transformations, the 
off-diagonal terms in this matrix must be set to zero. $\phi_P$ must therefore 
be set to one of

\begin{equation}
\phi_P = -\phi_C, \pi-\phi_C
\end{equation}
and we choose the former. The Clifford algebra to be used is now
\begin{equation}
\gamma_A = \{I, \gamma_P, \gamma_W, \gamma_{45}, \gamma_\mu, \gamma_{\mu P}, \gamma_{\mu W}, \gamma_{\mu 45} \}
\end{equation}
\begin{displaymath}
C = \gamma_2 \mbox{e}^{i\phi_C \gamma_{45}},\: \Pi = \gamma_1 \gamma_P.
\end{displaymath}

With the Clifford algebra and chiral multiplets established we now look for a 
general multiplet. This is very similar to the standard general multiplet in 
3+1 dimensions except for an extra scalar field $K$, which makes up the bosonic
degree of freedom lost when the vector field is taken from 3+1 to 2+1 
dimensions. Our general multiplet $V$ is defined by the following fields and 
transformations:
\begin{equation}
\begin{array}{ccl}
\delta C & = & \bar{\zeta} \gamma_W \chi \\
\delta \chi & = & (M + i\gamma_W N ) \zeta + i\gamma^\mu (A_\mu + i\gamma_W
\partial_\mu C) \zeta - \gamma_P K \zeta \\
\delta M & = & \bar{\zeta} ( \not\!\partial \chi + i\lambda ) \\
\delta N & = & i\bar{\zeta} \gamma_W ( \not\!\partial \chi + i\lambda ) \\
\delta A_\mu & = & \bar{\zeta} \gamma_\mu \lambda - i\bar{\zeta} \partial_\mu
\chi \\
\delta K & = & -i\bar{\zeta} \gamma_P \lambda \\
\delta \lambda & = & \frac{1}{2} (\gamma^\nu \gamma^\mu - \gamma^\mu \gamma^\nu )\partial_\mu A_\nu \zeta + i\gamma_W D \zeta + i\gamma_P \not\!\partial K \zeta \\
\delta D & = & i\bar{\zeta} \gamma_W \! \! \not\!\partial \lambda.
\end{array}
\end{equation}

\setcounter{equation}{0}
\section{Solving the DSEs of SQED3} \label{DSE}

Repeating the method of Wess and Zumino \cite{wz} yields the SQED3
lagrangian
\begin{equation} \label{lagrang} \begin{array}{l}
L = |f|^2 + |g|^2+ |\partial_\mu a|^2 + |\partial_\mu b|^2
- \bar{\psi} \not \! \partial \psi \\ \\
-m(a^*f + af^* + b^*g + bg^* + i\bar{\psi} \psi)\\ \\

- ieA^\mu(a^\ast \stackrel{\leftrightarrow}{\partial}_\mu a
+ b^\ast \stackrel{\leftrightarrow}{\partial}_\mu b
+ \bar{\psi} \gamma_\mu \psi)

+ e K\bar{\psi} \gamma_P \psi \\ \\

- e[\bar{\lambda}(a^\ast + i\gamma_W b^\ast)\psi 
- \bar{\psi}(a + i\gamma_W b) \lambda] \\ \\

+ ieD(a^\ast b - a b^\ast)
-e^2 (K^2 - A_\mu A^\mu) (|a|^2 + |b|^2) \\ \\

-\frac{1}{4}F^{\mu \nu}F_{\mu \nu} - 
     \frac{1}{2}\bar{\lambda} \not \! \partial \lambda
+ \frac{1}{2} \partial^\mu K \partial_\mu K + \frac{1}{2} D^2
\end{array} \end{equation}
which becomes that found by Pisarski \cite{Pski} 
by dimensional reduction of SQED4 when $\phi_C$ is set
to zero and the scalar fields are trivially redefined. 

The chiral limit is defined by taking $m\rightarrow 0$.  In this limit the 
bare lagrangian is invariant with respect to a global $U(2)$ symmetry 
generated by $I$, $\gamma_4$, $\gamma_5$ and $\gamma_{45}$. In the 
non-supersymmetric chiral theory, spontaneous mass generation leads to a 
nonperturbative breaking of the chiral generators $\gamma_4$ and $\gamma_5$.  
Here we shall explore the possibility of chiral symmetry breaking in 
SQED3 by considering the chiral multiplet propagator DSEs.  We shall use 
Witten's result that supersymmetry remains unbroken~\cite{witt,Pski}.  

\begin{figure}[h]
\vskip 0in
\psfig{figure=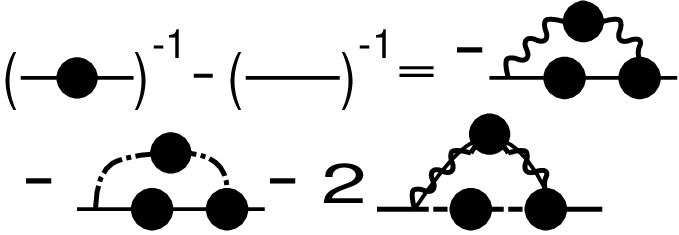}
\caption{The Dyson-Schwinger equation for the electron propagator in
SQED3. \label{fig:DSE}}
\end{figure}

The electron propagator DSE in SQED3 is more complicated 
than that in the non-supersymmetric version since the electron interacts 
not only with the photon, but also with the photino, 
\begin{figure}[h]
\vskip 0in
\psfig{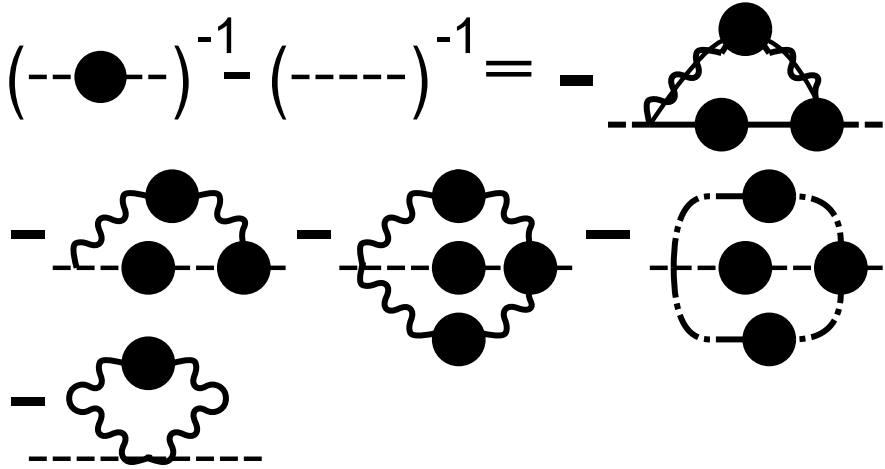}
\caption{The Dyson-Schwinger equation for the $a$ and $b$ propagator in
SQED3. \label{fig:DSEab}}
\end{figure}
the $K$ and its own
superpartners $a$ and $b$. The DSE for the electron is illustrated in
Fig.\ref{fig:DSE},
where a solid line indicates an electron propagator,  dashed line an $a$ or
$b$ propagator, a wavy line a photon propagator, an alternating dash-and-dot
line a $K$ propagator and a wave with a solid line through it a photino
propagator. 

The DSE for the scalar partners is shown in Fig.\ref{fig:DSEab}. It would 
appear {\it prima facie} that we have two coupled DSEs 
to solve, the second considerably more complicated than the 
first. We are saved from this effort however, by the existence of 
supersymmetric Ward 
Identities (WI) \cite{iz}. The supersymmetric WI relating the scalar two point 
functions to the electron propagator is
\begin{equation} \label{SUSY_WI}
\langle \psi \bar{\psi} \rangle  = 
i\langle a^* f\rangle  - i\not \! p \langle a^* a\rangle  =
i\langle b^* g\rangle  - i\not \! p \langle b^* b\rangle .
\end{equation}
Substituting in the fermion propagator Ansatz
\begin{equation} \label{fermiprop}
S(p) \equiv \langle \psi \bar{\psi} \rangle  = 
\frac{-i}{\not \! p A(p^2) + B(p^2)}
\end{equation}
gives the scalar propagator 
\begin{equation} 
\label{boseprop} 
D(p^2) \equiv \langle a^* a\rangle  = \langle b^* b\rangle  = 
\frac{A(p^2)}{p^2 A^2 (p^2) - B^2 (p^2)}, 
\end{equation}
and the $a$ to $f$ amplitude 
\begin{equation}
\label{atofprop}
\langle a^* f\rangle  = \langle b^* g\rangle  = 
\frac{B(p^2)}{p^2 A^2 (p^2) - B^2 (p^2)}
= \frac{B(p^2)}{A(p^2)} D(p^2).  
\end{equation} 
If nonperturbative chiral symmetry breaking is allowed for, the two point 
function $\langle a^* f\rangle $ is potentially non-zero in the chiral 
limit even though perturbation theory predicts it should be identically 
zero.  

\begin{figure}[ht] 
\psfig{figure=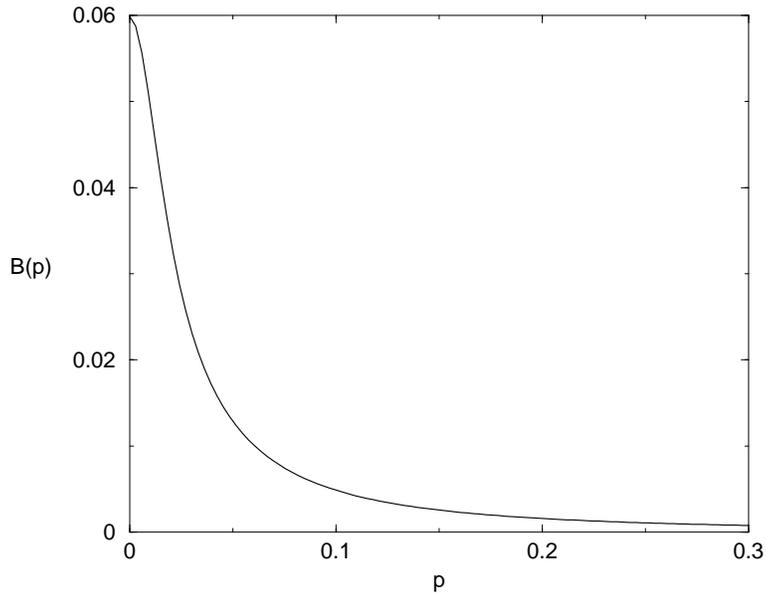,height=80 mm}
\caption{The chirally asymmetric solution for the propagator function 
$B(p^2)$ in Feynman gauge, $\xi = 1$.  If chiral symmetry is not broken, 
$B$ is identically zero.  \label{fig:Bofp}}
\end{figure}

\begin{figure}[ht]
\psfig{figure=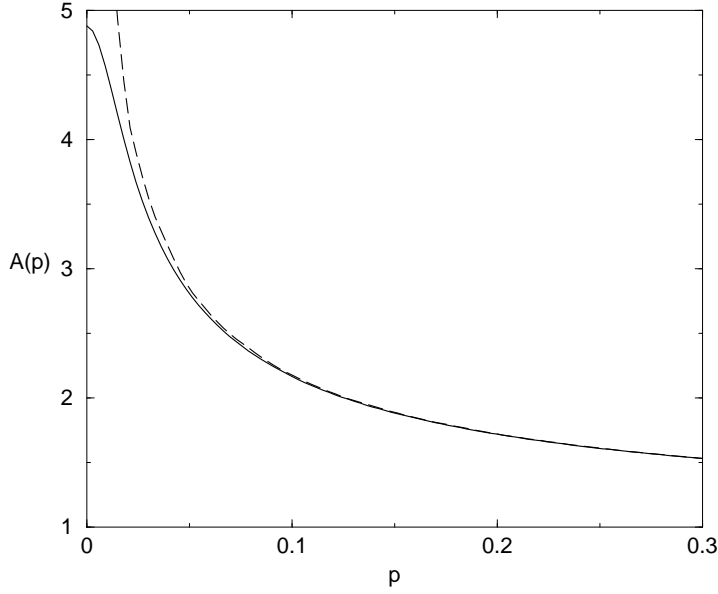,height=80 mm}
\caption{The chirally asymmetric (solid curve) and symmetric (dashed
curve) solutions for the propagator function $A(p^2)$ in Feynman gauge, 
$\xi = 1$.  \label{fig:Aofp}}
\end{figure}

\begin{figure}[ht]
\psfig{figure=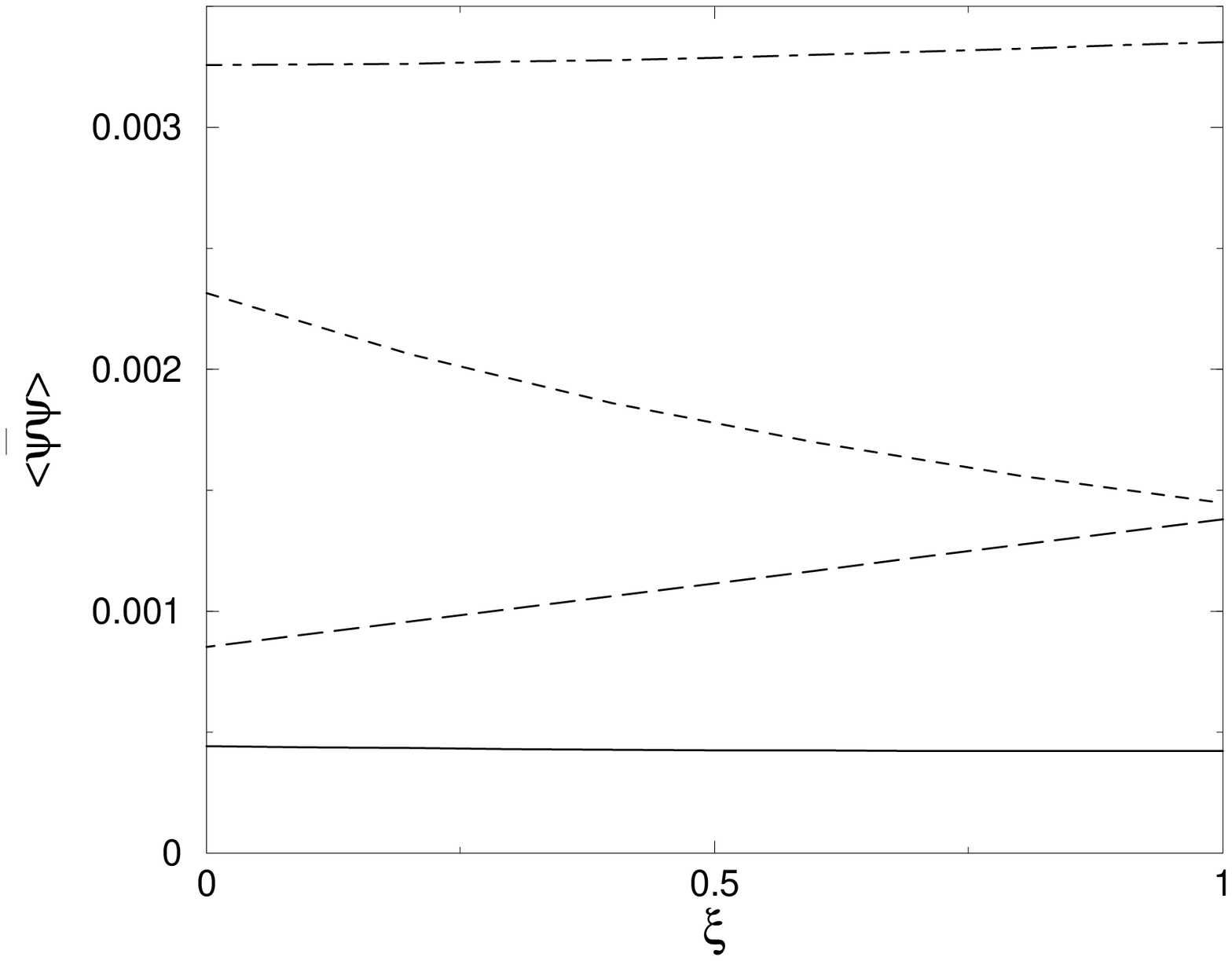,height=80 mm}
\caption{The chiral condensate of quenched SQED3 (solid curve) and 
QED3 (short dashes) in the bare vertex approximation.  Also plotted 
is the chiral condensate of quenched QED3 using the minimal Ball-Chiu 
Ansatz for the fermion-photon vertex (dashed-dot curve) and the 
chiral condensate of quenched SQED3 using the partial approximation 
to the Ball-Chiu Ansatz described in the text (long dashes).
  \label{fig:fig5}}
\end{figure}

Clearly it suffices to solve only the fermion DSE to determine the scalar
functions $A(p^2)$ and $B(p^2)$. To do the calculations we perform a Wick
rotation to Euclidean space.  Our Euclidean conventions are the same 
as those set out in Appendix B of reference~\cite{burd}.  Spacelike momenta 
satisfy $p^2 > 0$, and the fermion and boson propagators take the form 
\begin{eqnarray}
S(p) & = & \frac{1}{i\gamma\cdot p A(p^2) + B(p^2)}, \\
D(p) & = & \frac{A(p^2}{p^2 A(p^2) + B^2(p^2)},
\end{eqnarray}
respectively. We use the chirally quenched 
approximation~\cite{Love} which is known to be compliant
with $U(1)$ gauge invariance, that is, we take the propagators of the 
photon and its superpartners to be bare. The class
of covariant gauges, for which the photon propagator is given by
\begin{equation}
D_{\mu\nu}(k)=(\delta_{\mu\nu}-k_\mu k_\nu /k^2)\frac{1}{k^2} 
+ \xi \frac{k_\mu k_\nu}{k^4}
\end{equation}
will be considered.  The corresponding $K$ and photino
propagators are
\begin{eqnarray}
D_K(k) & = & \frac{1}{k^2}, \\
D_\lambda (k) & = &\frac{1}{i\gamma\cdot k},
\end{eqnarray}
respectively. We also employ the bare 
vertex or rainbow approximation
\begin{eqnarray}
\Gamma_\mu(q,p) & = & ie\gamma_\mu, \\
\Gamma_P(q,p) & = & -ie\gamma_P, \\
\Gamma_{\bar{\lambda} \psi} (q,p) & = & -ie, \\
\Gamma_{\bar{\psi} \lambda} (q,p) & = & ie,
\end{eqnarray}
for the Euclidean electron--photon, electron-$K$ and electron-$a$-photino 
vertices respectively.
The Euclidean space DSE with these approximations reduces
after angular integration to the following coupled integral
equations for $B(p^2)$ and $A(p^2)$:
\begin{eqnarray}
B(p^2) & = & (\xi + 3)\frac{e^2}{4\pi^2 p} \int_0^\infty dq
\frac{qB(q^2)}{q^2 A^2(q^2) + B(q^2)}\ln\left|\frac{p+q}{p-q}\right| 
\label{SDEB} \\
&& \nonumber\\
A(p^2) & = & (\xi -1)\frac{e^2}{4\pi^2 p^2} \int_0^\infty dq
\frac{qA(q^2)}{q^2 A^2(q^2) + B(q^2)}
\left(\frac{p^2 + q^2}{2p}\ln\left|\frac{p+q}{p-q}\right| - q\right) 
\label{SDEA} \nonumber \\
&& +\frac{e^2}{2\pi^2 p} \int_0^\infty dq 
\frac{qA(q^2)}{q^2 A^2(q^2) + B(q^2)}\ln\left|\frac{p+q}{p-q}\right| + 1
\end{eqnarray}

We solve equations (\ref{SDEB}) and (\ref{SDEA}) numerically using the standard
iterative procedure
introduced by Applequist et al.\cite{A86}. The functions $A(p^2)$ and $B(p^2)$
are defined
on a non-uniform grid of fifty-one points concentrated at small momenta 
where the function varies more rapidly. The integrand is interpolated
using a cubic spline using an ultraviolet cut-off of $p=1000e^2$.

We show in figures \ref{fig:Bofp} and \ref{fig:Aofp} both the chirally 
symmetric and asymmetric solutions to the massless ($m=0$), Feynman 
gauge ($\xi=1$) electron DSE of Fig.\ref{fig:DSE}.  The results are plotted 
in units with $e^2 = 1$.  It is apparent in figure
\ref{fig:Bofp} that mass generation is remarkably suppressed and both graphs
descend steeply to the values assumed when there is no dressing. We
have tested the convergence of our numerical iteration procedure by varying
the initial guess for $A$ and $B$ and the ultra-violet cutoff. These changes
had no significant effect on the solution obtained.

The DSE is the first functional derivative with the respect to the electron
propagator of the CJT effective potential~\cite{cjt} and its solutions are
therefore stationary points. To ascertain if the chirally asymmetric solution 
to the DSE is dynamically favoured we make use of the CJT 
effective potential which, when evaluated at a stationary point, is given 
by~\cite{StamRob}
\begin{eqnarray} 
\label{effpot}
V[S,D]&=&\int \frac{d^3p}{(2\pi)^3}\left\{\mbox{tr ln}[1 - \Sigma_S(p) S(p)]
  + \mbox{$\frac{1}{2}$} \mbox{tr}[\Sigma_S(p) S(p)]\right\} \nonumber \\ 
&& -2\int \frac{d^3p}{(2\pi)^3}\left\{\mbox{tr ln}[1 - \Sigma_D(p) D(p)]
  + \mbox{$\frac{1}{2}$} \mbox{tr}[\Sigma_D(p) D(p)]\right\},
\end{eqnarray}
where $\Sigma_S(p) = S(p)^{-1} - S_{\rm bare}^{-1}(p)$ and 
$\Sigma_D(p) = D(p)^{-1} - D_{\rm bare}^{-1}(p)$.  
This formula neglects the contribution from the photon and its superpartners
since it is calculated in the quenched approximation. Note that there is
another term in Eq.~(\ref{effpot}), given by
\begin{equation} \label{tadpole}
\int \frac{d^3p}{(2\pi)^3} \frac{d^3q}{(2\pi)^3} \left[D_{\mu \mu}(q) 
+ D_K(q)\right]D(p),
\end{equation}
corresponding to the vacuum graphs
giving rise to the tadpole contributions in the boson SDE. As we have
written it, Eq.~(\ref{effpot}) only gives half of these particular
diagrams. However in the quenched approximation Eq.~(\ref{tadpole})
is zero by dimensional reduction~\cite{itz}.

In Table~\ref{tab:effpot} we list the difference 
\begin{eqnarray} 
 V_{\rm A}[S,D] - V_{\rm S}[S,D] & = &
  2\int\frac{d^3p}{(2\pi)^3}\left\{\ln\left[\frac{p^2 A_{\rm S}(p^2)^2}
    {p^2 A_{\rm A}(p^2)^2 + B_{\rm A}(p^2)^2}\right] \right. \nonumber \\
 & - &\left.
 \frac{p^2 A_{\rm A}(p^2)}{p^2 A_{\rm A}(p^2)^2 + B_{\rm A}(p^2)^2} 
   + \frac{1}{A_{\rm S}(p^2)} \right\},
\end{eqnarray} 
between the effective potentials calculated for the chirally asymmetric and 
symmetric solutions for a range of values of the gauge parameter $\xi$.  
Our calculation confirms that the asymmetric solution is dynamically 
favoured in the nonsupersymmetric case.  For SQED3, however, we find 
that chiral symmetry is {\em not} spontaneously broken, at least in the 
quenched, rainbow approximation.  

\begin{table}[t]
\caption{The difference $V_{\rm A}[S,D] - V_{\rm S}[S,D]$ between the CJT 
effective potential of the chirally asymmetric solution and the chirally 
symmetric solution for QED3 and SQED3 at different values of 
the gauge parameter $\xi$.\label{tab:effpot}}
\vspace{0.2cm}
\begin{center}
\footnotesize
\begin{tabular}{|c|c|c|}
\hline
\raisebox{0pt}[13pt][7pt]{$\xi$} &
\raisebox{0pt}[13pt][7pt]{SQED3} &
\raisebox{0pt}[13pt][7pt]{QED3}\\
\hline
\raisebox{0pt}[13pt][7pt]{0} &
\raisebox{0pt}[13pt][7pt]{$4.62 \times 10^{-3}$}&
\raisebox{0pt}[13pt][7pt]{$-1.32 \times 10^{-5}$}\\
\hline
\raisebox{0pt}[13pt][7pt]{0.5} &
\raisebox{0pt}[13pt][7pt]{$4.29 \times 10^{-3}$} &
\raisebox{0pt}[13pt][7pt]{$-6.02 \times 10^{-6}$}\\
\hline
\raisebox{0pt}[13pt][7pt]{1} &
\raisebox{0pt}[13pt][7pt]{$4.07 \times 10^{-3}$} &
\raisebox{0pt}[13pt][7pt]{$-3.44 \times 10^{-6}$}\\
\hline
\end{tabular}
\end{center}
\end{table}

It is clear from Table~\ref{tab:effpot} that the calculated effective 
potential is dependent on the gauge parameter $\xi$.  This is an artifact 
of the rainbow approximation which, for dressed electron propagators, 
violates the $U(1)$ Ward-Takahashi identity (WTI).  Another useful indicator 
of gauge symmetry breaking is the $U(1)$ invariant chiral condensate 
\begin{equation}
\langle \bar{\psi} \psi \rangle = {\rm tr}S(x=0) = 
\frac{2}{\pi^2} \int_0^\infty dp \frac{p^2 B(p^2)}
                        {p^2 A^2(p^2) + B^2(p^2)},  
\end{equation}
calculated in the asymmetric phase.  
In Figure~\ref{fig:fig5} we plot the chiral condensate for quenched rainbow 
SQED3 and QED3.  While the calculated condensate for SQED3 is surprisingly 
insensitive to the choice of gauge, the condensate for QED3 is, as expected, 
strongly gauge dependent.  

For QED3 the invariance of the chiral 
condensate can be considerably improved by replacing the bare vertex 
with the minimal Ball-Chiu vertex Ansatz~\cite{BC}, which is specifically 
designed to respect the $U(1)$ WTI and be free of kinematic singularities.  
For comparison, the QED3 condensate obtained in this way in 
reference~\cite{BR91} is plotted in Figure~\ref{fig:fig5}.  Here we 
attempt a similar substitution for SQED3.  Note that the 
photon's supersymmetric partners $K$ and $\lambda$ are completely invariant 
under a gauge transformation and their vertices are not directly constrained 
by the WTI. Compliance with the WTI can therefore be achieved by replacing 
the bare photon-fermion vertex 
with the minimal Ball-Chiu Ansatz whilst the remaining    
vertices are kept bare. This method incurs the penalty of breaking 
supersymmetry.  The resulting chiral condensate is plotted in 
Figure~\ref{fig:fig5}.  Suprisingly the variation of the 
condensate with respect to the gauge parameter was found to be an order 
of magnitude greater than in the bare case.  
We attribute this to the violation of supersymmetry and conclude that 
any attempt to improve the vertex must remain supersymmetric.

\setcounter{equation}{0}
\section{Conclusions}

We have used the method of Wess and Zumino~\cite{wz} to obtain a lagrangian
for SQED3 with four-component fermions.  The lagrangian derived in 
this way agrees with 
that obtained by dimensional reduction~\cite{Pski} except that our form
retains an angular degree of freedom associated with the definition of 
parity and charge conjugation in $(2+1)$ dimensions~\cite{burd}.  
This is effected by taking suitable 
linear combinations of the Dirac matrices $\gamma_4$ and $\gamma_5$. It 
was found that supersymmetry constrains the angles $\phi_P$ and $\phi_C$ 
associated with parity and charge conjugation transformations in 
QED3~\cite{burd} to be related by $\phi_P = -\phi_C$ or $ \pi-\phi_C$.

The DSE of SQED3 was solved numerically in quenched, rainbow approximation 
after using supersymmetric WIs to relate scalar and fermion propagators 
within the chiral multiplet.  Application of the CJT effective
potential revealed that the chirally symmetric solution is dynamically 
preferred, in contrast to the situation in QED3.  We conclude that chiral 
symmetry is not broken in SQED3 in the quenched, rainbow approximation.  

Although the rainbow approximation violates the U(1) gauge symmetry, we find 
the chiral condensate to be relatively insensitive to the choice of gauge 
fixing parameter.  We believe this to be fortuitous however, since 
replacing the bare vertex with an Ansatz which respects the $U(1)$ WTI 
but not the supersymmetric WIs produces a chiral condensate which is 
significantly gauge dependent.  Work on developing a vertex Ansatz 
which respects both the $U(1)$ gauge and supersymmetric WIs is currently 
in progress.  

\section*{Acknowledgements}
The authors gratefully acknowledge stimulating discussions with V. Miransky. We
are also grateful to the Special Research Centre for the Subatomic Structure
of Matter, Adelaide, for hosting the Workshop on Nonperturbative Methods in
Quantum Field Theory which made these discussions possible.


\end{document}